\documentstyle{rspublic}

\def\ltsima{$\; \buildrel < \over \sim \;$}
\def\gtsima{$\; \buildrel > \over \sim \;$}
\def\simlt{\lower.5ex\hbox{\ltsima}}
\def\simgt{\lower.5ex\hbox{\gtsima}}

\def\hmpc{h^{-1}{\rm Mpc}}
\def\etal{{\it et al.}\rm}

\begin{document}

\title[Ly $\alpha$ absorption]{\bf Ly$\alpha$ Absorption
Systems and the Intergalactic Medium}

\author[G. Efstathiou, J. Schaye, T. Theuns]{G. Efstathiou$^{\rm 1}$, 
J. Schaye$^{\rm 1}$,
T. Theuns$^{\rm 1, 2}$}

\affiliation{1. Institute of Astronomy, Madingley Road,\\ 
Cambridge, CB3 OHA, UK\\
2. Max-Plank-Institut f\"ur Astrophysik, Postfach 1523, \\
85740 Garching, Germany}

\label{firstpage}

\maketitle

\begin{abstract}{Intergalactic medium, galaxy formation, dark matter}
The last few years have seen a dramatic improvement in our
understanding of the origin of Lyman $\alpha$ absorption
systems. Hydrodynamic numerical simulations of cold dark matter
dominated universes have shown that the many properties of the Lyman
$\alpha$ absorption systems can be explained by a photoionized,
space-filling, intergalactic medium. Lyman $\alpha$ lines offer
 promising probes of the photoionizing background, the
amplitude of the mass fluctuations at high redshift and the
evolution of the equation of state of the intergalactic medium.
\end{abstract}

\section{Introduction}

The existence of a forest of absorption lines blueward of the
Ly$\alpha$ emission line in quasar spectra has been known for over 30
years (Bahcall and Salpeter 1965; Lynds 1971). These lines arise from
Ly$\alpha$ absorption by neutral hydrogen from intervening structure
along the line-of-sight. Early theoretical models interpreted this
structure as absorption caused by discrete gas clouds in the
intergalactic medium (IGM), either pressure confined by a hot IGM
(Sargent \etal\ 1980; Ostriker and Ikeuchi 1983) or confined by the
gravity of dark matter `mini-halos' ({\it e.g.}  Rees 1986). Over the
last few years our understanding of the Ly$\alpha$ forest has
undergone a transformation for at least two reasons. Firstly,
observations with the Keck telescope have produced almost noise-free
spectra of quasars at high spectral resolution over the redshift range
$2 \simlt z \simlt 4$. The exquisite quality of Keck spectra has
allowed observers to resolve Ly$\alpha$ absorption lines at low column
densities ($ \sim 10^{12.5}\; {\rm cm}^{-2}$) and to study their
evolution. Secondly, hydrodynamic numerical simulations of structure
formation in cold dark matter (CDM) universes with high spatial
resolution are now possible and have proved remarkably successful in
reproducing many observed properties of the Ly$\alpha$ forest (Cen
\etal\ 1994; Zhang , Anninos and Norman 1995, 1997; Miralda-Escud\'e
\etal\ 1996; Hernquist \etal\ 1996, Theuns \etal\ 1998a). These
simulations have shown that most of the Ly$\alpha$ lines at column
densities $\simlt 10^{14.5} {\rm cm}^{-2}$ arise from modest
fluctuations in the baryon density in a space filling photoionized
IGM, rather than from distinct clouds. The properties of the
Ly$\alpha$ lines can therefore be used to probe the structure and
thermal history of the diffuse IGM and of the background UV radiation
that determines its ionization state. The key characteristics of the
numerical simulations are described in the next Section. Section 3
summarizes a number of results from these simulations and describes
how the Ly$\alpha$ lines can be used to study the IGM. In this paper
we discuss only the properties of the low column density Ly$\alpha$
lines. For a discussion of damped Ly$\alpha$ systems and metal lines
see Pettini's contribution to these proceedings. For a recent review
of observations and theoretical models of Ly$\alpha$ absorption lines
see Rauch (1998).

\section{Numerical Simulations of the IGM}

\begin{figure}
\vspace{7.5cm} 
\includegraphics{pghaardt.ps} 
\includegraphics{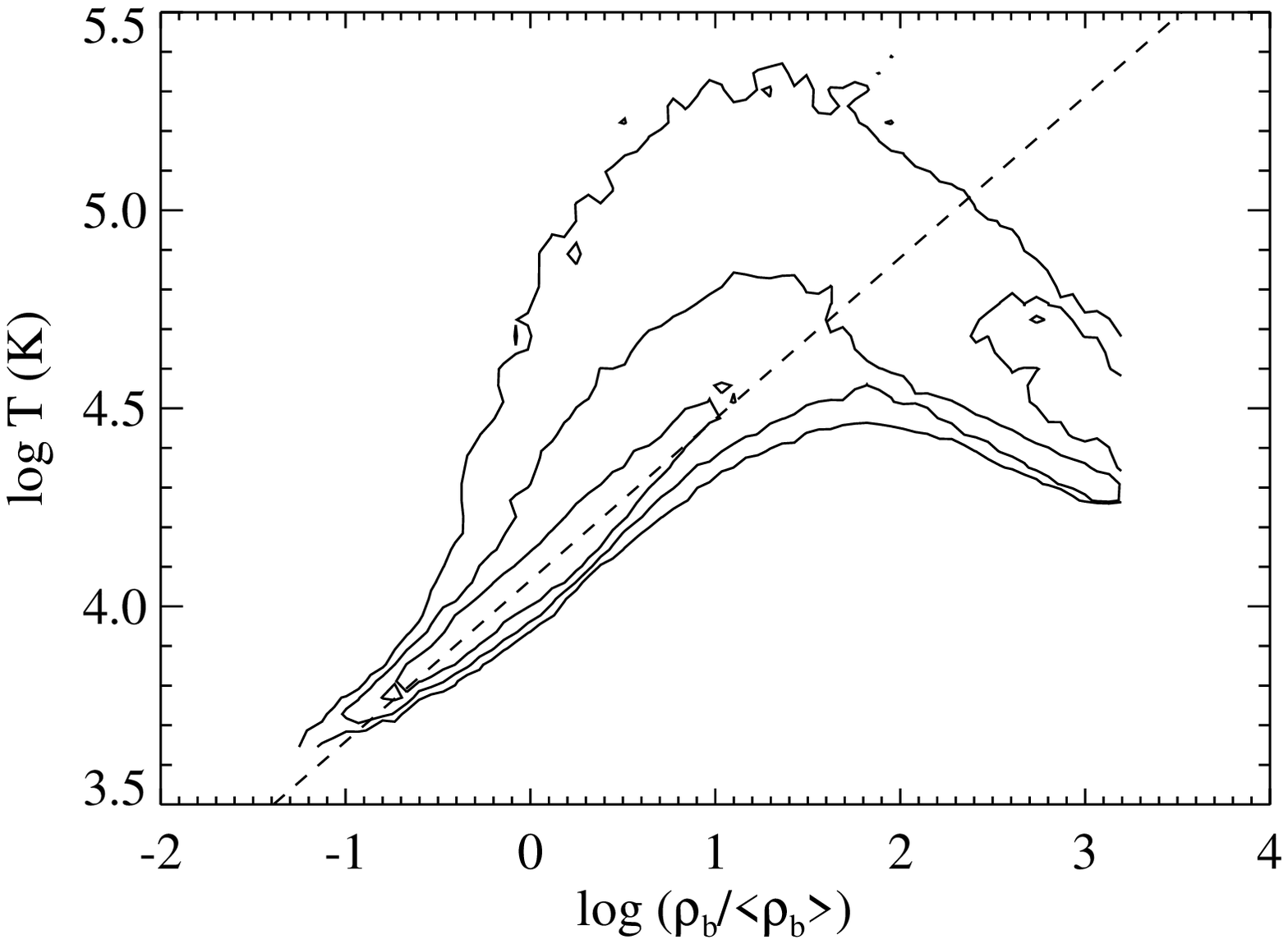}
\caption{The figure to the left shows
photoionization rates for hydrogen and helium computed
in the model of Haardt and Madau (1996). The figure to the right shows
the mass weighted distribution of fluid elements in the temperature-density 
plane at $z=3$ for our reference CDM model. The number density
of fluid elements increases by an order of magnitude with each contour
level. Most of the gas obeys a well defined equation of state shown by
the dashed line.}
\end{figure}

The simplest cosmological hydrodynamical simulations follow the
evolution of (optically thin) gas and dark matter assuming a uniform
photoionizing background. The simulation is therefore specified by:

\smallskip

\noindent
$\bullet$ parameters defining the cosmological model and its matter
content ({\it e.g.} $\Omega_m$, $\Omega_b$, $\Omega_\Lambda$, $H_0
\equiv 100 h {\rm km}{\rm s}^{-1}{\rm Mpc}^{-1}$);

\smallskip

\noindent
$\bullet$ the amplitude and spectral shape of the mass fluctuations;

\smallskip

\noindent
$\bullet$ a model for the background UV flux as a function of redshift.

\smallskip

Here we will review the evolution of CDM universes with initially
scale-invariant adiabatic fluctuations. The linear power spectrum for
these models (in the limit of small baryon content) is given by
Bardeen \etal\ (1986). We adopt a reference model (Model S) with
parameters $\Omega_m = 1$, $\Omega_\Lambda = 0$, $h=0.5$ and $\Omega_b
= 0.05$. The physical density in baryons in this model is $\omega_b
\equiv \Omega_bh^2 = 0.0125$. The model is normalized so that the rms
density fluctuations in spheres of radius $8\hmpc$ is $\sigma_8 =
0.7$. The photoionizing background radiation is assumed to originate
from quasars according to the model of Haardt and Madau (1996,
herafter HM). The photoionization rates 
for hydrogen and helium in this model are plotted in the left hand
panel of Figure 1 as a function of redshift. With these
photoionization rates, and assuming a uniform IGM, hydrogen is
reionized at a redshift of $z \approx 6$ and HeII is reionized at $z
\approx 4.5$.

The right hand panel of Figure 1 shows the distribution of gas
elements in the temperature-density plane in a numerical simulation of
our reference model S at a redshift of $z=3$. There is a plume of
shock heated gas extending to temperatures $\simgt 10^5$K, but most of
the gas has a low overdensity and follows a power law-like `equation
of state'
\begin{equation}
T = T_0 \left (\frac{\rho_b}{\bar \rho_b} \right ) ^{\gamma-1},  \label{1}
\end{equation}
(Hui and Gnedin 1997). At times long after reionization, the diffuse IGM will
settle into a state in which adiabatic cooling is balanced by
photoheating. The temperature of the IGM will therefore tend towards 
\begin{eqnarray}
T \approx 3.2 \times 10^4\;{\rm K}\; \left[ \frac{\Omega_b h (1 + \delta)
(1 + z)^3}{ (2 + \alpha) E(z)} \right ]^{1/1.76}, \label{2}  \\
E(z) = \frac{H_0}{H(z)} =  [ \Omega_m (1 + z)^3 + \Omega_k (1 + z)^2
+ \Omega_\Lambda]^{1/2} \nonumber
\end{eqnarray} 
where we have assumed a power-law photoionizing background 
\begin{equation}
J_\nu = J_{\nu_L} \left ( \frac{\nu}{\nu_L} \right )^{-\alpha}.
\end{equation}
In equation (2.2) $\Omega_k = 1 - \Omega_m
- \Omega_\Lambda$ and the exponent $1/1.76$  arise from the temperature
dependence of the HII and HeIII recombination coefficients.

Figure 2 shows the spatial distribution of the gas in the simulation
at $z=3$. The upper figure shows the gas with temperature $T>10^{5}$K.
This hot gas fills a small fraction of the volume and is located in
the dense knots and filaments corresponding to collapsed
structures. In contrast,  the cool gas with $T < 10^5$K (shown in the
lower figure) fills most of the computational volume. It is this diffuse
low density gas which we believe accounts for the vast majority of the
observed Ly$\alpha$ lines. Furthermore, because most of this gas is
at low overdensities, $\delta \simlt 10$, it is relatively easy to
model numerically. By investigating the Ly$\alpha$ forest we can
therfore hope to learn about the properties of the diffuse IGM, which at
typical quasar redshifts of $ z\sim 2 -4$ contains most of the baryonic
matter in the Universe. In particular:

\noindent
$\bullet$ the optical depth of HI and HeII absorption can set
constraints on the baryonic density of the Universe and on the
evolution, amplitude and spectrum of the photoionizing background;

\noindent
$\bullet$ the fluctuating optical depth of HI absorption can be used
to construct the power spectrum of the matter fluctuations at redshifts
$z \sim 2-4$;

\noindent
$\bullet$ the Ly$\alpha$ absorption line-widths can be used to infer
the temperature and equation of state of the diffuse IGM and its
evolution.

 These topics will be discussed in more detail in the next section.

\vfill\eject

\begin{figure}[t]
\vspace*{2 truecm}
\centering{\texttt{fig2a.gif}}\\
\vspace*{4 truecm}
\centering{\texttt{fig2b.gif}}\\
\vspace*{2 truecm}
%\vspace{18.5 truecm}
%\special{psfile=cube.Tgt1e5K.ps hscale=68 vscale=68 angle=0 hoffset=  0
% voffset= 115}
%\special{psfile=cube.Tlt1e5K.ps hscale=68 vscale=68 angle=0 hoffset=  0
% voffset=-150}
\caption{The distribution of gas in a CDM simulation at $z=3$. The
upper figure shows shocked gas with a temperature greater than $10^5$K,
which is located in dense clusters and filaments. The lower panel
shows gas with a temperature $< 10^5$K.}
\end{figure}
%\noindent
%Figure 2. The distribution of gas in a CDM simulation at $z=3$. The
%upper figure shows shocked gas with a temperature greater than $10^5$K,
%which is located in dense clusters and filaments. The lower panel
%shows gas with a temperature $< 10^5$K.
%\addtocounter{figure}{1}

\noindent

\begin{figure}[t]
\vspace{10.5cm}  
\includegraphics{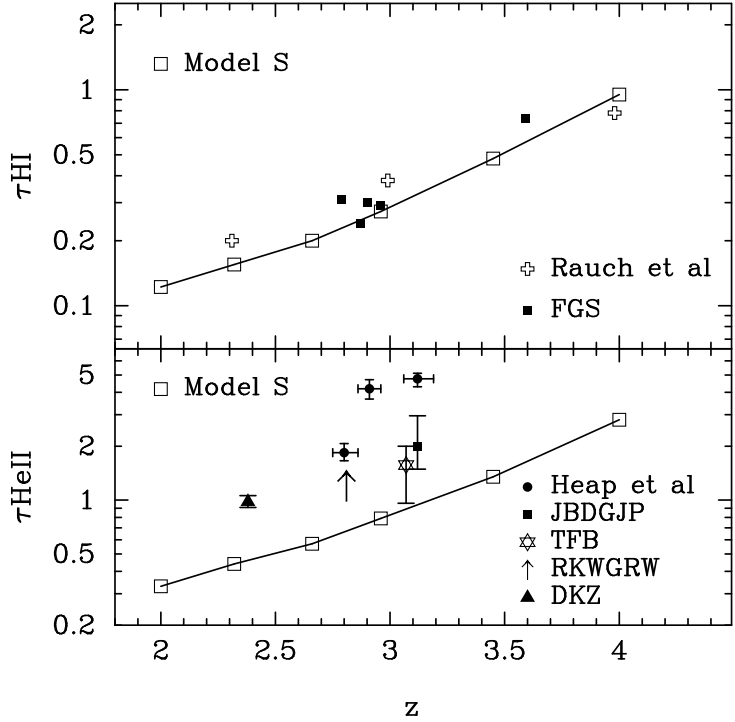}
\caption{The mean optical depth as a function of redshift for HI and
HeII absorption. Our reference model S (using half the amplitude of the
HM photoionizing background) is plotted as the open squares
joined by the solid line. This matches the observed HI optical depth
but fails to match observations of the HeII optical depth. 
The observational data
plotted in the figure are as follows: Rauch \etal\ 1997; Fardal \etal\
1998 (FGS); Davidsen \etal\ 1996 (DKZ); Jakobsen \etal\ 1997 (JBDGJP);
Reimers \etal\ 1997 (RKWGRW); Tytler, Fan and Burles 1995 (TFB); Heap \etal\
2000.}
\end{figure}

\section{The Ly $\alpha$ forest as a probe of the IGM}

\subsection{Mean optical depth of HI and HeII absorption}

The optical depth for HI Ly$\alpha$ absorption from an IGM in ionization
equilibrium with density $\rho_b$ is given by
\begin{equation}
\tau(z) = 6.5\times 10^{-4} \left ( \frac{\omega_b}{0.019} \right)^2
\left ( \frac{h}{0.65} \right )^{-1} \frac{(1+z)^6}{E(z)}
\frac{T_4^{-0.76}}{(\Gamma_{\rm HI} /10^{-12} {\rm s}^{-1}) } \left (
\frac{\rho_b(z)} {\bar \rho_b(z)} \right )^2, \label{3.1}
\end{equation}
({\it e.g.} Peebles 1993 \S23), where $\Gamma_{\rm HI}$ is the
photoionization rate for HI and $T_4$ is the temperature of the IGM in
units of $10^4$\;K. Variations in $\rho_b(z)$ along the line-of-sight
will produce a `fluctuating Gunn-Peterson' effect. An observed
absorption line spectrum can therefore be inverted to infer the
clustering of the baryon distribution as pioneered by Croft \etal\
(1998, see Section 3c below).

The mean HI and HeII optical depths as a function of redshift are
plotted in Figure 3 for our fiducial model S. The open squares show
the optical depths derived using half the amplitude of the HM
UV background. With this choice of photoionizing background, 
the mean HI optical depth of the simulation matches observations quite
well over the redshift range $2$--$4$. Evidently, the amplitude of the
photoionizing background can be balanced by variations in other
cosmological parameters according to equation (\ref{3.1}) to preserve
the match to observations. The simulations thefore imply that
\begin{equation}
 \left ( \frac{\omega_b}{0.0125} \right)^2 \left ( \frac{h}{0.5}
\right )^{-1} \Omega_m^{1/2} \left ( \frac{0.5 \Gamma_{\rm
HM}}{\Gamma_{\rm HI}} \right ) \left ( \frac{6 \times 10^3}{T} \right
)^{0.76} \approx 1. \label{Omb}
\end{equation}
This type of criterion has been used by Weinberg \etal\ (1997) to
set a crude lower limit to the baryon density. At redshifts $2$--$3$
one can be reasonably confident of the HM model of the photoionizing
background at around the Lyman edge, because the quasar luminosity
function is quite well determined at these redshifts (see Section 3 of
Madau's article in these proceedings). The HM model provides a lower
limit to the photoionizing flux because it ignores additional
photoionizing radiation from star formation. The temperature of a
photoionized IGM cannot exceed a few times $10^4$K, but its precise
value depends on the past thermal and ionization history of the
IGM.\footnote{Although a highly ionized IGM is in ionization
equilibrium, it is not in thermal equilibrium and so retains a memory
of the way in which it was heated.} Uncertainties in the temperature
of the IGM lead to an additional source of uncertainty in using
equation (\ref{Omb}) to derive a bound on the baryon
density. Nevertheless, equation (\ref{Omb}) implies $\omega_b \simgt
0.02 \Omega_m^{1/2}$, interestingly close to the baryon density of
$\omega_b = 0.019$ inferred from primorial nucleosynthesis and the
deuterium abundances measured in quasar spectra (Burles and Tytler
1998, Burles, Kirkman and Tytler 1999). The observed HI optical depth
therefore leads to a consistent picture in which most of the baryons
in the Universe at $z\sim 2$--$4$ belong to the diffuse photoionized
IGM.

The HeII optical depth is shown in the lower panel of Figure 3.  The
open squares show $\tau_{\rm HeII}$ computed from the simulation using
the same amplitude for the photoionizing background that provides a
good match to $\tau_{\rm HI}$. The results from the simulation lie
below the observations at all redshifts, suggesting that the
photoinizing background has a softer spectrum than computed by HM. In
photoionization equilibrium, the optical depth in HeII is related to
the optical depth in HI by $\tau_{\rm HeII}/\tau_{\rm HI} \propto
\Gamma_{\rm HI}/ \Gamma_{\rm HeII} \propto J_{\rm HI}/J_{\rm HeII}$,
and so $\tau_{\rm HeII}$ can be raised by softening the photoionizing
spectrum appropriately. By lowering $\Gamma_{\rm HeII}/\Gamma_{\rm
HI}$ by a factor of two compared to the HM model of Figure 1, model S
can match the observations at $z \simlt 2.8$, but cannot match the
high HeII optical depths at $z \simgt 2.9$ found in the recent
HST-STIS observations of the quasar Q0302-003 by Heap \etal\ (2000).

 One possible interpretation of these results is that the typical
quasar spectrum adopted by HM is too hard and that HeII reionization
is delayed until a redshift $z\approx 3$, corresponding to the abrupt
change in HeII opacity observed by Heap \etal\ (2000). Additional
arguments to support this picture are discussed in Section (3d). If
this interpretation is correct, then AGN and `mini-quasars' cannot
produce much photoionizing radiation capable of doubly-ionizing helium
at redshifts $z \simgt 3$ (see the article by Rees in this volume).

\subsection{Evolution of the column density distribution}

It has been known for many years that Ly$\alpha$ forest shows strong
cosmological evolution ({\it e.g.} Sargent \etal\ 1980). Over a
narrow range in redshift the evolution can be approximated by a power
law
\begin{equation}
        \frac{dN}{dz} \propto (1+z)^\epsilon, \label{evol}
\end{equation}
where $N$ is the number of lines above a threshold rest frame
equivalent width (typically $W>0.32 \AA$). From high resolution Keck
observations,  Kim \etal\ (1997) find $\epsilon = 2.78 \pm 0.71$ in
the redshift range $2 < z < 3.5$ . Williger \etal\ (1994) find that
the evolution is still stronger at higher redshifts with $\epsilon >
4$ at $z > 4$. In contrast, observations with the HST indicate much
weaker evolution at low redshifts with $\epsilon = 0.48 \pm 0.62$ for
$z < 1$ (Morris \etal\ 1991; Bahcall \etal\ 1991, 1993; Impey
\etal\ 1996).

\begin{figure}
\vspace{10cm}  
\includegraphics{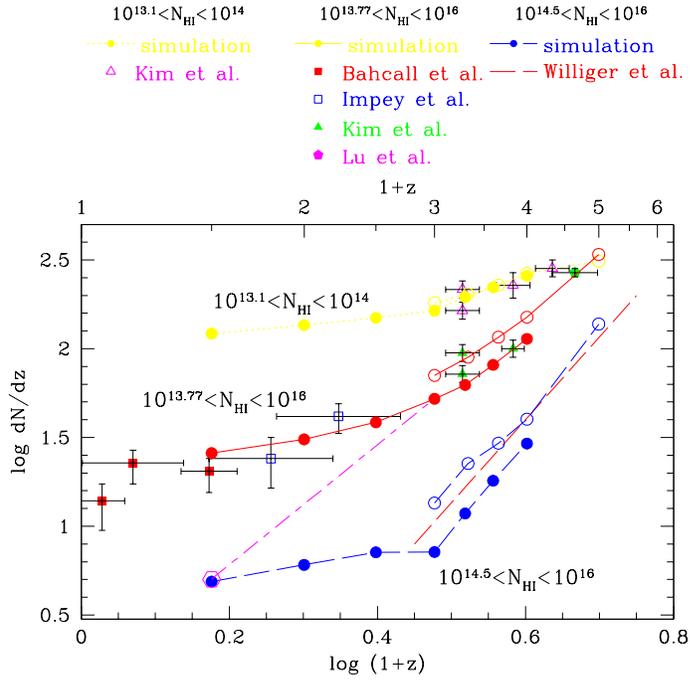}
\caption{Evolution of the number of lines within a given range of
column density from numerical simulations of Theuns \etal\ (1998b)
compared with observations.  Open and filled circles show simulation
results for the reference model S (described in the text) run at
different numerical resolutions. The large open pentagon shows a
reanalysis of the simulation at $z=0.5$ but imposing the photoionizing
background appropriate to $z=2$. The observational data points are as
follows: Kim \etal\ 1997 (open and filled triangles); Bahcall
\etal\ 1993 (filled squares); Impey \etal\ 1996 (open squares);
Lu \etal\ 1996 (filled pentagon); Williger \etal\ 1994 (long
dashed line).}
\end{figure}

The evolution of the Ly$\alpha$ forest, including the low rates of
evolution at redshifts $z \simlt 2$ can be reproduced quite simply in
CDM models (Theuns, Leonard and Efstathiou 1998b; Dav\'e \etal\ 1999). This is
illustrated in Figure 4 which shows the evolution of the number 
of Ly$\alpha$ lines within a given range of column density for
our reference model S. (The lines are identified by fitting Voigt
profiles to simulated Keck spectra using the line-fitting program
VPFIT, Webb 1987).  As in Section (3a) we adopt the HM
model of the photoionizing background with an amplitude divided by a
factor of $2$ to match the observed optical depth in HI
absorption. This model reproduces the observed column density
distribution accurately over the column density range $10^{12.5}\;{\rm
cm}^{-2} \simlt N_{\rm HI} \simlt 10^{15}\; {\rm cm}^{-2}$ (see Figure
2 of Theuns \etal\ 1998b) and, as Figure 4 shows, also reproduces
the observed rates of evolution as a function of column density. In
particular, the decrease in the rate of evolution of the Ly$\alpha$
lines found from HST observations arises from the steep decline in the
photoionizing background at $z \simlt 2$ caused by the rapid drop in
quasar numbers at low redshift.

\subsection{Reconstruction of the matter power spectrum}

Equations (2.1) and (3.1) can be combined to write the observed
transmitted flux in terms of fluctuations in the baryon density,
\begin{equation}
F = {\rm exp} \left [ - A (\rho_b/\bar \rho_b)^\beta \right], \qquad 
\beta \approx (2.76 - 0.76 \gamma).
\end{equation}
Croft \etal\ (1998) have used this relation to infer the 1-dimensional
power spectrum $P_{1D}(k)$ of the baryon fluctuations, 
from which the three-dimensional power
spectrum can be recovered by differentiation,
\begin{equation}
P(k) = -\frac{2 \pi}{k} \frac{d} {dk} P_{1D}(k).
\end{equation}
Croft \etal\ (1998) calibrate the amplitude of the matter power
spectrum by comparing against numerical simulations. The procedure is
not completely straightforward and we refer the reader to Croft
\etal\ (1998) for details.  By testing their inversion algorithm
against numerical simulations these authors find that the amplitude
and shape of the underlying matter power spectrum can be recovered
accurately and that the recovery is insensitive to uncertainties in
the equation of state. A variant of this technique, that incorporates
a correction for the distortion of the clustering pattern by peculiar
velocities,  is described by Nusser and Haehnelt (1999).

Croft \etal\ (1999) describe an application of their method to a
sample of $19$ quasar spectra spanning the redshift range
$2.08$--$3.23$. They recover $P(k)$ at $z \approx 2.5$ over the
(comoving) wavenumber range $2\pi/k \sim 2$--$12\; \hmpc$ and find
that it is well fitted by a power law $P(k) \propto k^n$ with $n =
-2.25 \pm 0.28$, consistent with what is expected from CDM
models. This important result is the first attempted determination of
the matter power spectrum at high redshift. The amplitude of the
inferred power spectrum is high compared to that expected for spatially
flat CDM models with $\Omega_m =1$ normalized to reproduce the
abundance of rich clusters at the present day.  The best fitting CDM
models have $\Omega_m + 0.2\Omega_\Lambda \approx 0.46$ (Weinberg
\etal\ 1999; Phillips \etal\ 2000), in agreement with the
parameters $\Omega_m \approx 0.3$, $\Omega_\Lambda \approx 0.7$
derived from combining observations of anisotropies in the cosmic
microwave background radiation and distant Type Ia supernovae ({\it
e.g.} Efstathiou \etal\ 1999 and references therein). Constraints
on neutrino masses from estimates of $P(k)$ at high redshift are
discussed by Croft, Hu and Dav\'e (1999).

\subsection{Widths of the Ly$\alpha$ lines}

The widths of the Ly$\alpha$ lines are usually characterised by the
broadening parameter $b$ determined by fitting Voigt profiles. If the
line width is caused by thermal broadening, the parameter $b$ is
related to the temperature of the gas by
\begin{equation}
 b =  \left ( \frac{2kT}{m_p}\right )^{1/2} = 
12.8\; T_4^{1/2} \;{\rm km}\;{\rm s}^{-1}. \label{broad}
\end{equation}
In reality, a number of mechanisms in addition to thermal broadening
contribute to the widths of the lines, {\it e.g.} the differential
Hubble flow across the absorbing region and the smoothing of small
scale fluctuations by gas pressure (Bryan \etal\ 1999; Theuns, Schaye and
Haenhelt 2000).

\begin{figure}
\vspace{6cm}  
\includegraphics{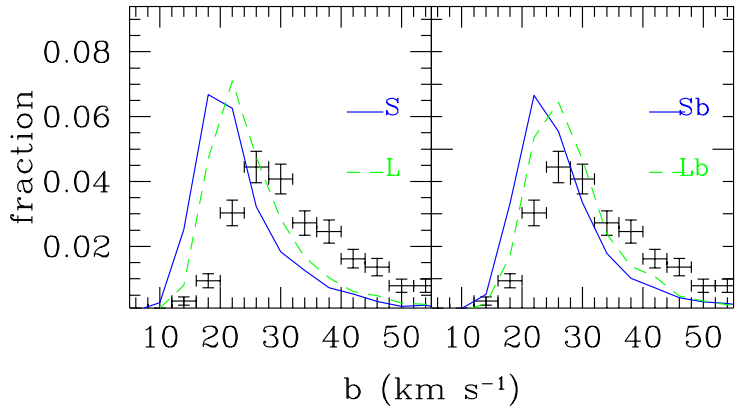}
\caption{The figure to the left shows the distribution of line widths
at $z=3$ derived from simulations of the reference model S and for a
spatially flat $\Lambda$ dominated universe (model L) with
$\Omega_\Lambda=0.7$ and identical photoionizing background. Both of
these models have a baryon density of $\omega_b = 0.0125$. The figure
to the right shows models with $\omega_b = 0.025$ and with double the
amplitude of the photoionizing background so that the optical depth of
HI absorption is almost unchanged. The error bars show observational results
from Hu \etal\ 1995.}
\end{figure}

The first sets of simulations of the Ly$\alpha$ forest in CDM models
appeared to show good agreement with the observed $b$-parameter
distributions. However, subsequent simulations with higher spatial
resolution produced a larger fraction of narrower lines than observed
(Theuns \etal\ 1998a; Bryan \etal\ 1999). This is illustrated in
Figure 5 which shows observations of the $b$ distribution at $z\approx
3$ compared to the results from numerical simulations. The figure to
the left shows the distribution derived by fitting Voigt profiles to
simulated spectra from reference model S (solid line). The lines in
this model are clearly too narrow, suggesting that the temperature of
the IGM is too low.  From the assymptotic relation (2.2) one can see
that the temperature of the IGM can be raised by increasing the baryon
density and by lowering $\Omega_m$ (the assymptotic temperature is
extremely insentive to the amplitude of the photoionizing background
long after reionization). The right hand panel shows the effect of
increasing the baryon density to $\omega_b = 0.025$.  The dashed lines
in each panel show the effect of lowering $\Omega_m$ and introducing a
cosmological constant so that the universe remains spatially
flat. These variations in cosmological parameters can go some way to
resolving the conflict with observations (Theuns \etal\ 1999), but
cannot provide an exact match unless the baryon density is much higher
than the value favoured from primordial nucleosythesis.  This suggests
that we are missing a significant heating source of the IGM. A number
of mechanisms have been suggested that might boost the temperature of
the IGM, {\it e.g.}  photoelectric heating of dust grains (Nath, Sethi
and Shchekinov 1999) and Compton heating by the hard X-ray background
(Madau and Efstathiou 1999). However, the most plausible explanation
(Abel and Haenhelt 1999) is that the simulations underestimate the
temperature at $z\sim 3$ because they assume an optically thin IGM
(and also a uniform photoionizing background that has {\it already}
been reprocessed by Lyman $\alpha$ absorbing clouds, Haardt and Madau
1996). This is inconsistent and the simulations should properly
include the effects of radiative transfer while the medium is still
optically thick prior to complete reionization, because in this regime
every photoionizing photon is absorbed and contributes to heating the
IGM. Abel and Haehnelt estimate that correct inclusion of radiative
transfer during the epoch of HeII reionization might increase the
temperature of the IGM by a factor of $\sim 2$, perhaps enough to
resolve the discrepancy with the observed $b$-parameter distributions
illustrated in Figure 5. The idea that the temperature of the IGM at
$z\sim 3$ is boosted by HeII reionization is supported by the analysis
of the equation of state of the IGM described in the next subsection.

\subsection{Constraining the equation of state of the IGM}

As we have mentioned, a number of physical mechanisms contribute to
the breadth of the $b$-distribution. However, the minimum line-width
is set by the temperature of the gas, which in turn depends on the
density ({\it cf.} Figure 1). By fitting the cut-off in the
$b$-distribution as a function of column density (the `$b(N)$'
distribution) one can therefore reconstruct the effective equation of
state of the IGM (Schaye, \etal\ 1999; Ricotti, Gnedin and Shull 2000;
Bryan and Machacek 2000).

\begin{figure}[t]
\vspace{5cm}  
\includegraphics{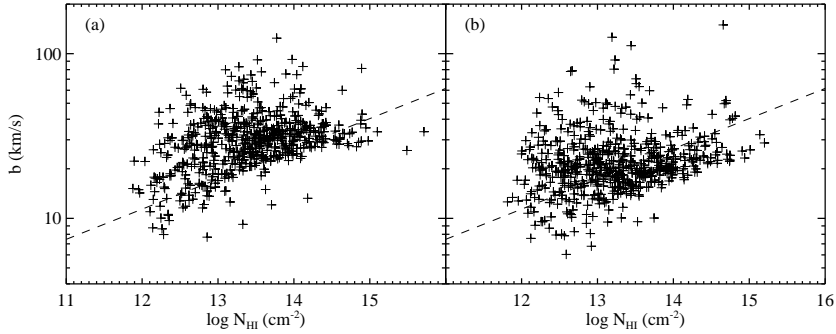}
\caption{The $b(N)$ distributions derived by applying the Voigt profile
fitting program VPFIT to 800 random lines of sight through two
cosmological simulations at $z=3$.  The figure to the left shows
results for a model with a hot IGM (see text). Only lines
with VPFIT errors $\Delta b/b < 0.5$ and $\Delta N_{HI}/N_{HI} < 0.5$
are plotted. The dashed line shows the best fit to the lower envelope
of the $b(N)$ distribution over the column density range $10^{12.5}
{\rm cm}^{-2} \le N_{HI} \le 10^{14.5} {\rm cm}^{-2}$ determined using
the algorithm of Schaye \etal\ (1999). This dashed line is reproduced
in the figure to the right together with the $b(N)$ distribution for
the colder reference model S.}
\end{figure}

The method is illustrated by Figure 6 which shows the $b(N)$
distributions derived by applying the VPFIT line-fitting program to
two cosmological simulations. Results for the standard reference model
S are plotted in Figure (6b). Figure 6(a) shows a hotter model, which
has the same parameters as model Lb plotted in Figure 5 but with the
HeI and HeII photoheating rates doubled over those of the HM model
(crudely representing `radiative transfer' effects during the
reionization of helium). The dashed line in the figure shows the best
fit to the lower envelope of the $b(N)$ distribution of Figure (6a)
determined by applying the iterative fitting algorithm of Schaye
\etal\ (1999). The same line is plotted in Figure (6b) and passes
almost through the middle of the $b(N)$ distribution of the colder
model. The lower envelope of the $b(N)$ distribution is clearly a
sensitive indicator of the characteristic temperature of the
Ly$\alpha$ clouds and can be accurately determined from fits to high
resolution quasar spectra.

What is more difficult is to convert the fit to the lower envelope of
the $b(N)$ distribution, $b = b_0(N/N_0)^{\Gamma-1}$, into the
parameters $T_0$ and $\gamma$ of the effective equation of state of
the IGM (equation 2.1).  This is done by calibrating $b_0$ and
$\Gamma$ against $T_0$ and $\gamma$ using numerical simulations with
different equations of state (see Schaye \etal\ 1999 for details).

The results of applying this technique to nine high-quality quasar
spectra spanning the redshift range $2.0 < z < 4.5$ are shown in
Figure 7 (Schaye \etal\ 2000). Except for the two lowest redshift
quasars, the Ly$\alpha$ forest spectra were divided in two to reduce
the effects of redshift evolution and signal-to-noise variations
across a single spectrum. From $z \sim 3.5$ to $z \sim 3$, the
inferred temperature at the mean density of the IGM, $T_0$, increases
and the gas is close to isothermal ($\gamma \sim 1$). This behaviour
differs markedly from that expected if helium were fully reionized at
higher redshift. For example, the solid lines show the evolution of
the equation of state in a simulation with the HM background.  In this
simulation, both hydrogen and helium are fully ionized by $z \approx
4.5$ and the temperature of the IGM declines slowly as the universe
expands tending to the assymptotic form of equation (2.2). This model
cannot account for the peak in the temperature at $z \sim 3$ inferred
from the observations. Instead, we associate the peak in $T_0$,  and the
low value of $\gamma$,  with reheating caused by the second reionization
of helium (HeII $\rightarrow$ HeIII). This interpretation is supported
by the abrupt change in HeII opacity at $z \sim 3$ from the
measurements of Heap \etal\ (2000) discussed in Section 3(a). It is
also consistent with evidence from the ratio SiIV/CIV that the
spectrum of the photoionizing background hardens abruptly at $z \sim
3$ (Songaila and Cowie 1996; Songaila 1998).\footnote{Note, however,
that Boksenberg, Sargent and Rauch (1998) find a more gradual change
of the SiIV/CIV ratio with redshift.}

The dashed lines in Figure 7 show a `designer model' with parameters
tuned to fit the observations. This simulation has a much softer UV
background at high redshift than the model shown by the solid line,
and consequently HeII reionizes at $z \sim 3.2$. In addition, the
photoheating rates have been enhanced during reionization to boost the
temperature of the IGM (again, crudely modelling the heating of optically
thick gas). This model is suggestive that the jump in
temperature at $z \sim 3$ may be associated with HeII reionization,
but clearly more realistic simulations that include radiative transfer
are required to determine the evolution of the equation of state
during the reionization phase more accurately. Further observations would
also be useful to assess how steeply the IGM temperature changes between
$z=4$ and $z=3$.

\begin{figure}[t]
\vspace{6cm}  
\includegraphics{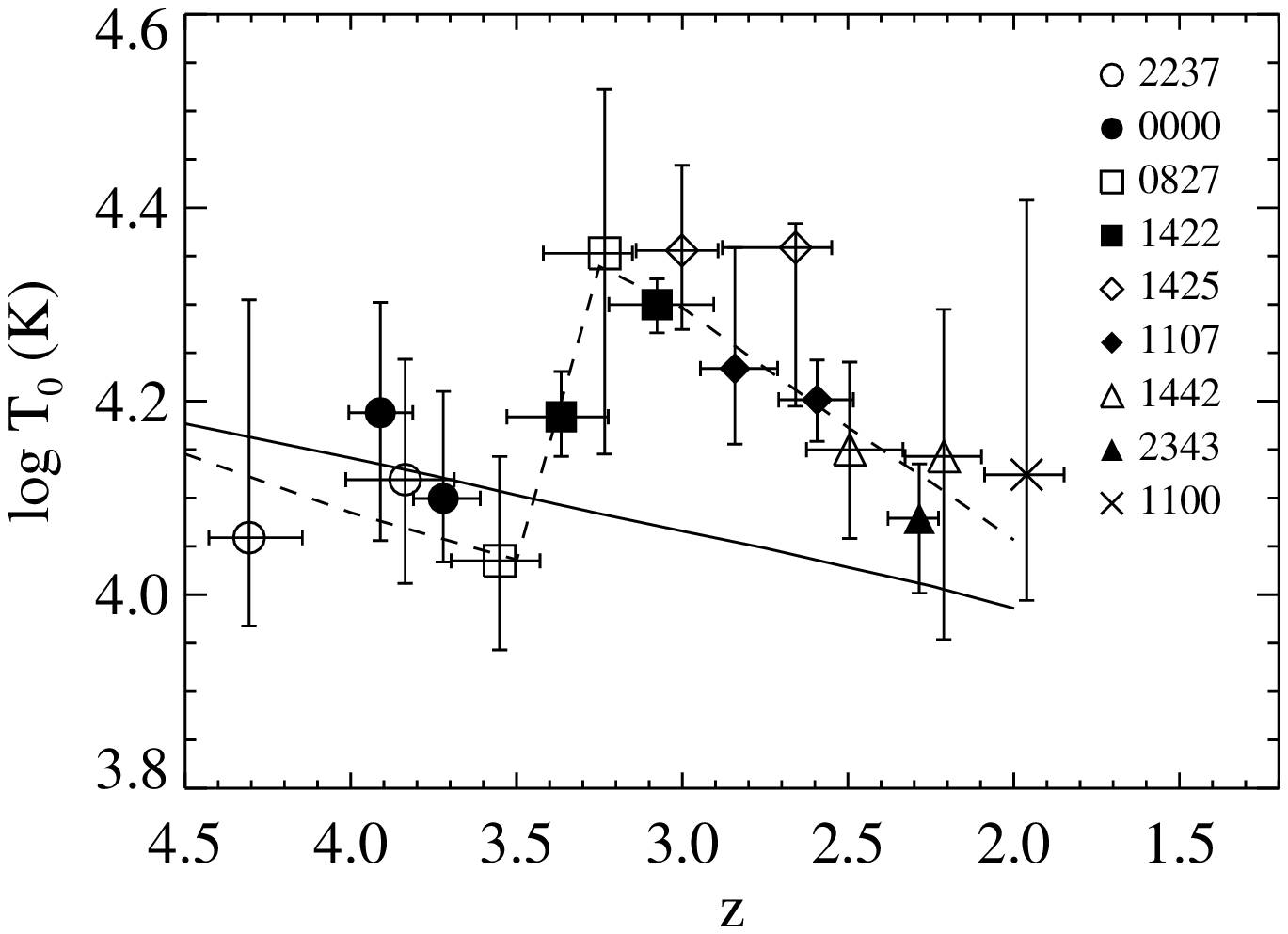}
\includegraphics{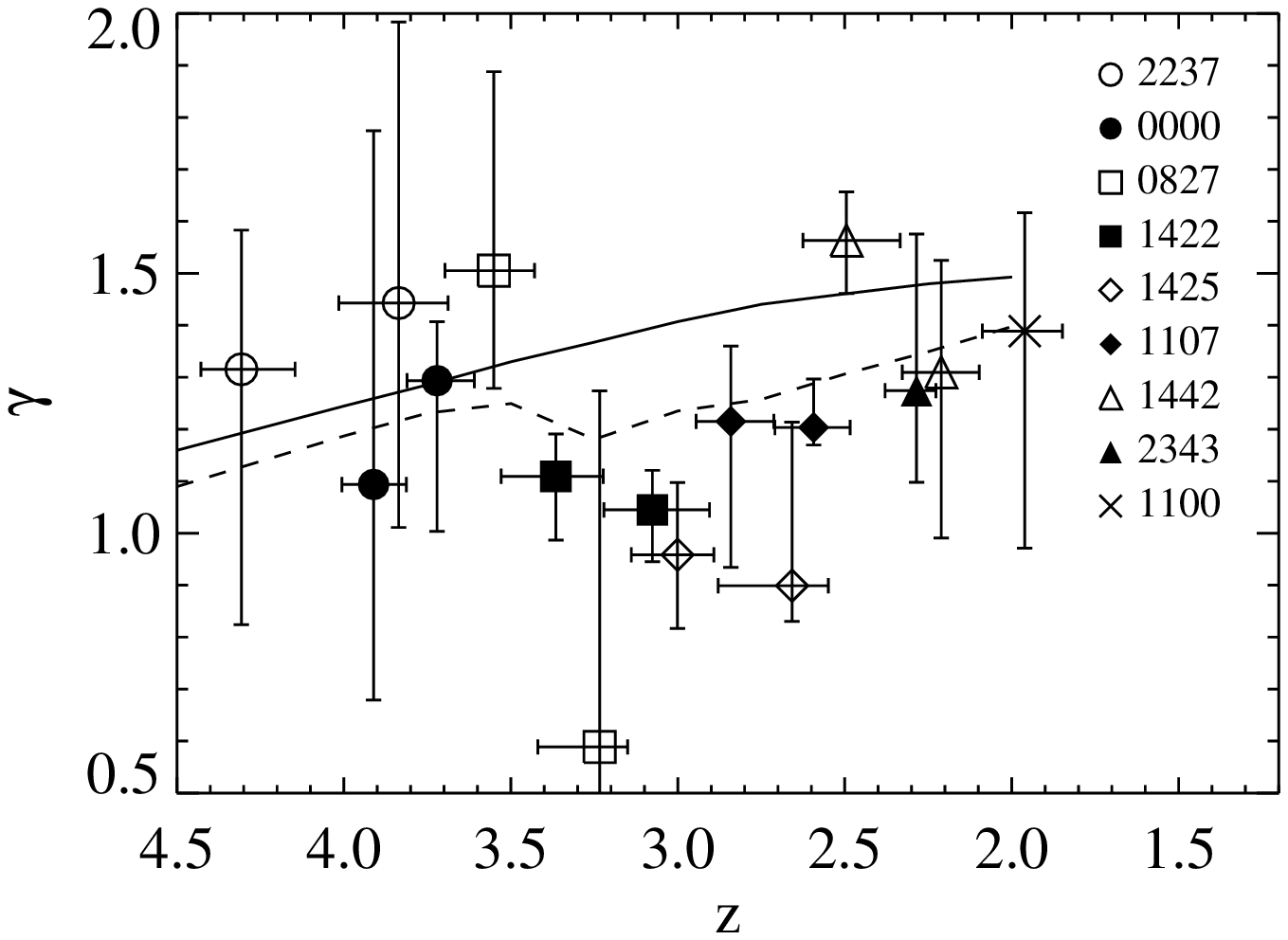}
\caption{The figure to the left shows the temperature at the mean
density as a function of (decreasing) redshift inferred from the lower
envelope in the $b(N)$ distribution determined from $9$ high
resolution spectra (Schaye \etal\ 2000). The figure to the right shows
the inferred slope of the equation of state as a function of
redshift. The horizontal error bars indicate the redshift interval
spanned by the absorption lines and the vertical error bars show
estimates of $1 \sigma$ errors. The lines show the evolution of the
equation of state in two numerical simulations as described in the
text. Different symbols correspond to different quasars.}
\end{figure}

\section{Summary and Outlook}

The work reviewed in this article provides a powerful case that the
Ly$\alpha$ forest arises from a space-filling, highly photoionized
diffuse IGM that contains most of the baryonic material in the
Universe at high redshift. This model is a natural outcome of CDM
theories of structure formation and can account for many observed
properties of the Ly$\alpha$ forest in quantitative detail. The
general features of the model thus seem to us to be reasonably secure.

However, a more detailed analysis of the thermal history of the IGM
requires simulations that incorporate radiative transfer and a model
for the spatial distribution of ionizing sources. Such calculations
are now being done (Abel, Norman and Madau 1999; Gnedin 2000;
Madau these proceedings) but the computational problems are
formidable. Some outstanding problems that deserve further attention
include:

\noindent
$\bullet$ detailed simulations of the inhomogeneous reionization of
hydrogen and helium;

\noindent
$\bullet$ extending the analysis of Ly$\alpha$ line widths to redshifts
$\simgt 4$, perhaps leading to constraints on the epoch of reionization
of hydrogen;

\noindent
$\bullet$ analysis of inhomogeneities in the temperature of the IGM.
Are there, for example, regions in the spectra of quasars in which 
Ly$\alpha$ line-widths are systematically broader or narrower than in
other regions?

\noindent
$\bullet$ further observations of absorption gaps in HeII Ly$\alpha$
absorption (reported by Heap \etal\  (2000) and others) and the
development of a model to understand their sizes;

\noindent
$\bullet$ searching for signatures of outflows around protogalaxies
in the Ly$\alpha$ forest;

\noindent
$\bullet$ determining the mean metallicity of the IGM and understanding
how the metals were transported from protogalaxies.

\begin{acknowledgements}
J. Schaye thanks the Isaac Newton Trust for financial support and
PPARC for the award of a studentship. We also thank Anthony Leonard,
Michael Rauch and Wal Sargent for their contributions to this work. 
\end{acknowledgements}

\end{document}